\documentclass[pra,twocolumn,a4paper,groupedaddress,showpacs,floatfix,aps,10pt]{revtex4-1}
\usepackage{graphicx,amsmath,amssymb,amsfonts,dsfont}
\newcommand{\mf}[1]{\boldsymbol{#1}}
\newcommand{\ket}[1]{\ensuremath{|#1\rangle}}
\newcommand{\mc}[1]{\ensuremath{\mathcal{#1}}}
\newcommand{\bra}[1]{\ensuremath{\langle #1 |}}
\newcommand{\braket}[2]{\ensuremath{\langle #1 | #2 \rangle}}

\newcommand{\mean}[1]{\ensuremath{ \langle #1  \rangle}}

\usepackage{color}

\begin{document}

\title{Quantum mechanical calculation of Rydberg-Rydberg autoionization rates}

\author{Martin Kiffner${}^{1,2}$}
\author{Davide Ceresoli${}^{3}$}
\author{Wenhui Li${}^{1,4}$}
\author{Dieter Jaksch${}^{2,1}$}

\affiliation{Centre for Quantum Technologies, National University of Singapore,
3 Science Drive 2, Singapore 117543${}^1$}
\affiliation{Clarendon Laboratory, University of Oxford, Parks Road, Oxford OX1
3PU, United Kingdom${}^2$}
\affiliation{Istituto di Scienze e Tecnologie Molecolari CNR, via Golgi 19,
20133 Milano, Italy${}^3$}
\affiliation{Department of Physics, National University of Singapore, 117542,
Singapore${}^4$}


%
%
%


\begin{abstract}
We present quantum mechanical calculations of   autoionization  rates for two Rubidium Rydberg
atoms with weakly overlapping electron clouds. We neglect exchange effects and consider tensor products of independent atom states 
forming an approximate basis of the two-electron state space. We consider  large sets of two-atom states with randomly chosen quantum numbers 
and find that the charge overlap between the two  Rydberg electrons allows one to characterise the 
magnitude of the autoionization   rates. If the electron clouds overlap by more than one percent,  
the autoionization   rates increase approximately exponentially with the charge overlap. This finding is independent of the energy of the initial state. 
\end{abstract}

\maketitle

\section{Introduction \label{intro}}
\begin{figure}[t!]
\begin{center}
\includegraphics[width=8.5cm]{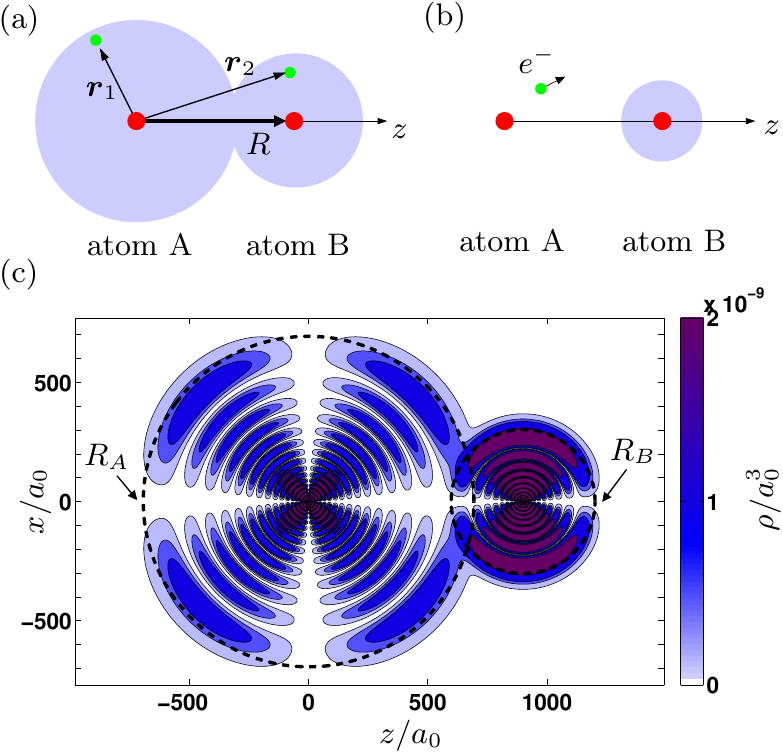}
\end{center}
\caption{\label{fig1}
(Color online)
(a) Two Rydberg atoms in different electronic states and with spatial separation $R$. The blue spheres indicate the size of the
Rydberg electron charge density cloud, and $\mf{r}_i$ are the electron coordinates. 
(b) Schematic illustration of the autoionization process. Atom 2 makes a transition into
a lower bound state and the electron of atom 1 is ejected into the continuum.
(c) Charge density $\rho$ for wave function $\Psi_{\text{M}}$ in Eq.~(\ref{state}) with $n_A=20$,  $l_A=2$, $m_A=-1$ 
and $n_B=15$,  $l_B=1$, $m_B=1$. The dashed sphere with radius $R_A$ ($R_B$) denotes the classical outer turning point of 
the electron of atom A (B). 
}
\end{figure}

Exciting  ultracold atoms to  Rydberg states~\cite{gallagher:ryd} with large principal quantum
number $n$  furnishes the atoms with extremely exaggerated properties. For example,  the size, interaction strength and 
polarizability increases by several orders of magnitude as compared to  ground state atoms. This feature 
allows one to study fundamental physical phenomena on completely new time and length scales and magnifies physical effects  such 
that they become experimentally accessible. For example, dipole-dipole interactions between ground state atoms are typically weak, but 
they are strong and long-ranged between Rydberg atoms such that  $\mu$m-sized molecules consisting of
two~\cite{boisseau:02,schwettmann:06,schwettmann:07,overstreet:09,samboy:11,samboy:11b,kiffner:12,kiffner:13}
and three~\cite{samboy:13,kiffner:13l,kiffner:14} atoms become possible. 
Moreover, dipole-dipole interactions between Rydberg atoms give rise to the blockade effect~\cite{urban:09,gaetan:09} and 
crystals of spatially ordered Rydberg excitations that were experimentally observed in ~\cite{schauss:12}. The modification of the quantum dynamics of 
Rydberg electrons due to their dipole-dipole interaction has been demonstrated in a recent experiment by Takei {\em et al.}~\cite{takei:15} 
via ultrafast pump-probe laser techniques.
In systems of dipole-dipole interacting Rydberg atoms the interatomic spacing is typically large  compared to the 
size of the Rydberg electron orbital. A fascinating prospect for future studies is the investigation of 
Rydberg systems with overlapping electron clouds. In this regime the exaggerated properties of Rydberg atoms would 
allow one to study the rich physics of electron-electron interactions 
on much more accessible  time and length scales  compared to conventional solid state systems. 
More specifically, the size of the valence electron orbital increases like $R_n = 4 n^2 a_0$ where $a_0$ is the Bohr radius and will
thus reach the typical separation between   atoms in optical lattices or tweezers for $n \gtrapprox 35$.  
Overlapping electron clouds could give rise to delocalized  electrons and correlated quantum many-body states via the strong Coulomb interaction 
between the electrons. However, these coherent processes compete with  autoionization and radiative decay processes enabled by the large number of empty 
orbitals below the Rydberg state. 
The first step in investigating the regime of Rydberg atoms with overlapping electron clouds is to characterize the time scales of 
the occurring physical processes. 
While radiative processes are well understood~\cite{beterov:09}, we here focus on 
autoionization of two neutral Rydberg atoms via the Penning effect~\cite{gallagher:ryd}  as shown in Fig.~\ref{fig1}. 
In this process  the energy for ionizing atom A is provided by a change in internal energy of atom B. 
Until now, quantum mechanical calculations of this process  are restricted  
to the dipole-dipole interaction regime of non-overlapping electron clouds~\cite{hahn:00,amthor:09} where the decay rates are 
negligibly small. On the contrary, calculations based on classical Hamilton equations~\cite{robicheaux:05} show that fast autoionization 
occurs for  atomic separations of the order of $R_n$ where the electron clouds start to overlap. This effect  
has been identified in~\cite{robicheaux:14} as a key factor for understanding the fast autoionization of a Rydberg gas observed in~\cite{tanner:08}. 
In order to determine the timescale of autoionization of Rydberg atoms with overlapping electron clouds, quantum mechanical calculations 
of the corresponding autoionization   rates are needed. 
However, a rigorous approach to this problem  is  extremely challenging  since it would involve finding  the highly excited two-electron eigenstates  of the system.

In order to estimate autoionization  rates of Rydberg atoms with overlapping electron clouds, we here present a simplified model 
and consider  two-atom states $\ket{\psi_M}$ with weakly overlapping electron clouds as shown in Fig.~\ref{fig1}. 
We assume that $\ket{\psi_M}$ is a tensor product of two generally different independent-atom orbitals and neglect exchange effects. 
In order to account for the fact that these states are not eigenstates of the system, we consider large sets of states $\ket{\psi_M}$ with different quantum numbers 
that could serve as an approximate basis of the true two-electron eigenstate. We evaluate the autoionization  rate of the states $\ket{\psi_M}$ 
quantum mechanically and show that the  charge overlap between the two atoms allows one to characterize the magnitude of the autoionization  rates. 
In the regime of very small charge overlap between the Rydberg orbitals, 
the autoionization rates are small and depend on the energy of the initial state. Moreover, 
the full interaction Hamiltonian can be approximated by its multipole expansion. 
On the contrary, above a certain threshold the multipole expansion becomes invalid and 
the autoionization  rates increase approximately exponentially with the charge overlap. 

Note that the autoionization mechanism between two Rydberg atoms considered here is related to autoionization processes 
in crystals and clusters that have been termed inter-atomic Auger decay~\cite{matthew:75} and more recently interatomic Coulombic decay (ICD)~\cite{cederbaum:97,averbukh:04,kuleff:10,ovcharenko:14}. 
In particular, the strong enhancement of autoionization rates through the overlap between electron orbitals in clusters was reported in~\cite{averbukh:04}, 
and ICD processes between several excited atoms in a cluster were studied in~\cite{kuleff:10,ovcharenko:14}. 

This paper is organized as follows. The system of interest and our model are described in Sec.~\ref{sys}.  We briefly outline the calculation 
of the autoionization  rates in Sec.~\ref{rates} and defer more technical details to Appendix~\ref{matrixE}. In order to account for many different initial states 
$\ket{\psi_M}$ we randomly select these states as described in Sec.~\ref{selection}. Finally, the  autoionization  rates of the randomly selected states are presented 
in Sec.~\ref{res} and a conclusion of our work is given in Sec.~\ref{conclusion}. 
%
%
%
\section{The system \label{sys}}
We consider two  Rydberg atoms as shown in Fig.~\ref{fig1}(a), where each atom is comprised of 
a singly-charged core and one valence electron. We assume that  the atoms are so cold that their positions do not change during the 
decay process. Atom A is centered
at the origin  and atom B is located at $\mf{R}=R\mf{e}_z$,
where $\mf{e}_z$ is the unit vector in $z$ direction and $R$ is the atomic
separation. 
The total Hamiltonian of  the two-atom system is
$ H = H_0 + V\,, $
where $H_0=H_0^{(A)} +H_0^{(B)}$ and $H_0^{(X)}$ is the Hamiltonian of Rydberg atom $X$. All interactions between  atom A
and atom B are described by  
\begin{align}
 V=& \frac{q^2}{4\pi\varepsilon_0}\left(\frac{1}{R}+\frac{1}{|\mf{\hat{r}}_1-\mf{\hat{r}}_2|} 
  -\frac{1}{|\mf{\hat{r}}_2|}-\frac{1}{|\mf{R}-\mf{\hat{r}}_1|}\right) \,,
 \label{V}
\end{align}
where $q$ is the elementary charge and $\mf{\hat{r}}_i$ the operator associated with the position of  electron $i$.  
The first term in Eq.~(\ref{V}) accounts for the repulsion of the two ion cores, the second is the electron-electron interaction and the third (fourth) term 
describes the interaction of electron 2 (1) with ion core A (B). 
The eigenstates of $H_0$ are 
\begin{align}
\ket{\Psi_{M}}=\ket{\psi_A,\psi_B},
\label{state}
\end{align}
where $\ket{\psi_A}$ and $\ket{\psi_B}$ are independent-atom Rydberg wavefunctions
centered at the origin and $\mf{R}$, respectively, 
\begin{align}
\psi_A(\mf{r}_1)=\psi_{n_A l_A m_A}(\mf{r}_1),\quad
\psi_B(\mf{r}_2)=\psi_{n_B l_B m_B}(\mf{r}_2-\mf{R}) .
\end{align}
We ignore the fine structure such that the wavefunctions $\psi_{nlm}(\mf{r})$ are characterised 
by the principal quantum number $n$, the orbital angular momentum quantum number $l$ and the azimuthal quantum number $m$. 
We generate the functions $\psi_{nlm}$ with energy $E_{n l} = -1/[n-\delta_{n}(l)]^2$ via the  
Numerov method, where $\delta_{n}(l)$ is the quantum defect~\cite{gallagher:ryd,zimmerman:79}. 
We choose  Rubidium 85 atoms which are a popular choice in recent Rydberg experiments and obtain the energies $E_{nl}$ (and hence the quantum defects) for $n\le 11$ from 
spectroscopic data reported in~\cite{luna:02}. For $n=11$, the non-zero quantum defects are $\delta_{11}(0)=3.134$, $\delta_{11}(1)=2.652$ and $\delta_{11}(2)=1.341$ which is consistent 
with the quantum defects provided in~\cite{li:03}. We ignore the weak dependence of $\delta_n(l)$ on the principal quantum number for $n>11$.  
Note that we order the quantum numbers in $\ket{\psi_M}$ such that $E_{n_A l_A}\ge E_{n_B l_B}$ by convention since the state obtained 
by interchanging $A$ and $B$ has the same autoionization  rate. 

An example for $\ket{\psi_M}$ is shown in Fig.~\ref{fig1}(c), where the  size of the electron cloud of atom $A$ ($B$)  is indicated 
by a sphere of radius $R_A$ ($R_B$), where $R_A$ ($R_B$) is the classical outer turning point,  
\begin{align}
 R_X= {n_X^*}^2 +n_X^*\sqrt{{n_X^*}^2-l_X(l_X+1)},
 \label{turning}
\end{align}
$n^*=[n-\delta_n(l)]$ is the effective quantum number and  $X\in\{A,B\}$. In order to quantify the overlap between the wavefunctions 
$\ket{\psi_A}$ and $\ket{\psi_B}$, we consider the amount of charge due to  $\ket{\psi_A}$ inside the sphere $V_B$ with radius $R_B$ around atom $B$, 
\begin{align}
\delta q_A(V_B) = q \int\limits_{V_B} |\psi_A|^2 d^3 r \,.
\end{align}
Similarly, 
\begin{align}
\delta q_B(V_A) =q \int\limits_{V_A} |\psi_B|^2 d^3 r 
\end{align}
is the amount of charge due to $\ket{\psi_B}$ inside the sphere $V_A$ with radius $R_A$ around atom $A$. 
A measure for the differential overlap between  $\ket{\psi_A}$  and $\ket{\psi_B}$  is then given by $\delta q/Q$, where $Q=2q$ is the total electron charge and 
\begin{align}
 \delta q = \delta q_A(V_B) + \delta q_B(V_A).
 \label{deltaq}
\end{align}
In the following we will consider only states with small overlap such that $\delta q/Q\ll 1$. 

Note that the physical wavefunction of the two-electron system should include spin degrees of freedom and be completely antisymmetric with respect to electron exchange. 
However, we find that the simplified state in Eq.~(\ref{state}) results in a good approximation of the autoionization  rate for weakly overlapping electron clouds as explained in Sec.~\ref{rates}.
\section{Autoionization  rates\label{rates}}
Next we outline the calculation of the autoionization  rate  for state $\ket{\psi_M}$ in Eq.~(\ref{state}). To this end, 
we consider the process shown in Fig.~\ref{fig1}(b) where atom $B$ makes a 
 transition to a lower bound state $\psi_b(\mf{r}_2)\equiv \psi_{n_b l_b m_b}(\mf{r}-\mf{R})$ with energy
$E_b\equiv E_{n_b l_b m_b}$,  and the other electron is ejected into the continuum. 
We model the  wavefunction of the  ejected electron with mass $m_e$ by energy-normalized Coulomb waves $\ket{\psi_{l m}^{E}}$~\cite{friedrich:tap,spencer:82} with 
angular momentum $l$, magnetic quantum number $m$ and energy $E$ obeying the generalized normalization relation 
\begin{align}
\braket{\psi_{l m}^{E}}{\psi_{l^{'} m^{'}}^{E^{'}}}=\delta(E-E^{'})\delta_{l l^{'}}\delta_{m m^{'}} \,. 
\end{align}
The Coulomb waves  are numerically generated by following the procedure described in~\cite{gallagher:ryd}. 
We calculate the autoionization  rate using Fermi's golden rule~\cite{tannoudji:api} and to first order in the interaction $V$. 
The decay rate $\Gamma_{M}^b$ for the process shown in Fig.~\ref{fig1}(b) is thus given by 
\begin{align}
\Gamma_{M}^b =  \frac{2\pi}{\hbar} \sum\limits_{l_k=0}^{\infty}\sum\limits_{m_k=-l_k}^{l_k}
|\bra{\psi_{l_k m_k}^{E_k},\psi_b}V\ket{\psi_M}|^2 \,,
\label{auger1}
\end{align}
where the energy $E_k=E_M-E_b$ of the Coulomb wave is fixed by energy conservation between the initial and final states and 
\begin{align}
E_M =\bra{\psi_M}H\ket{\psi_M} 
\label{energy} 
\end{align}
is the expectation value of the total Hamiltonian $H$ in the initial state $\ket{\psi_M}$. Note that $E_M$ differs by at most  $2\%$ from 
the unperturbed value $\bra{\psi_M}H_0\ket{\psi_M} =E_{n_A l_A}+E_{n_B l_B}$ for all states considered in Sec.~\ref{selection}. This is consistent with 
the wavefunctions comprising the initial state being only weakly perturbed by the electron-electron interaction in the overlap region. 

A more rigorous calculation with a fully antisymmetric initial state would result in two Coulomb matrix elements in Eq.~(\ref{auger1})  that are termed the direct and the 
exchange term~\cite{feibelman:77}. The single matrix element in  Eq.~(\ref{auger1}) corresponds to the direct term, and the exchange term is absent since our initial state $\ket{\psi_M}$ in Eq.~(\ref{state}) 
is a simple product state. The exchange term  depends on the overlap between the single-electron orbitals and 
decreases exponentially with increasing distance $R$~\cite{averbukh:04}. Since we are considering only weakly overlapping electron clouds we expect 
that exchange effects are small and hence the expression in Eq.~(\ref{auger1}) should be a good approximation for the autoionization  rate. 
%
\begin{figure}[t!]
\begin{center}
\includegraphics[width=8.5cm]{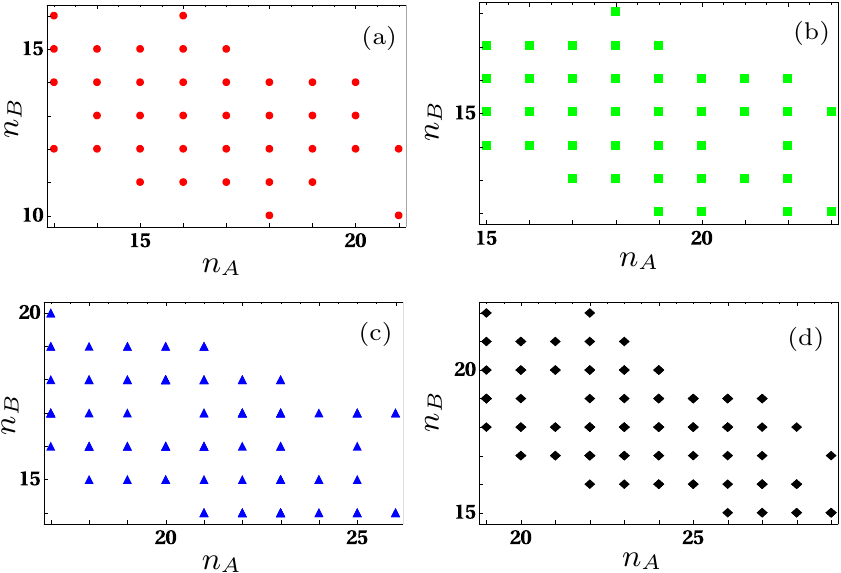}
\end{center}
\caption{\label{fig2}
(Color online)
Distribution of the principal quantum numbers $n_A$ and $n_B$ in the randomly chosen sets of states $\mc{S}_i$.  
(a) Red dots correspond to set $\mc{S}_1$. 
(b) Green squares show set $\mc{S}_2$. 
(c) Blue triangles are for set $\mc{S}_3$. 
(d) Black diamonds correspond to set $\mc{S}_4$.
}
\end{figure}

%
The decay rate $\Gamma_{M}^b$ in Eq.~(\ref{auger1}) accounts for all
processes where atom B makes a transition into the  bound state $\ket{\psi_b}$ and atom A is
ionized. In addition, we consider also the  autoionization process with rate $\tilde{\Gamma}_{M}^b$ 
where atom A makes a transition to  $\ket{\psi_b}$ and atom B is ionized. The full decay rate 
is then obtained by adding $\Gamma_{M}^b$  and $\tilde{\Gamma}_{M}^b$  and summing over all bound states, 
\begin{align}
 \Gamma_{M} 
  = \sum\limits_{\stackrel{b\ \text{with}}{E_b\le E_{M}}}\left(\Gamma_{M}^b + \tilde{\Gamma}_{M}^b\right).
 \label{auger2}
\end{align}
We numerically evaluate  Eq.~(\ref{auger2}) by restricting  the sum over bound states 
to those with $n_b\ge 0.2(n_A +n_B)$. This is justified since the contribution of lower-lying bound states is negligible. 
The  evaluation of the matrix element in Eq.~(\ref{auger2}) is described in detail in Appendix~\ref{matrixE}.  
In short, we expand all involved wavefunctions and the interaction Hamiltonian $V$ in Eq.~(\ref{V}) in terms of spherical harmonics and
limit the integration region to the volume where $\ket{\psi_{A,B}}$ both
take on non-negligible values. We restrict the maximum angular momentum in the expansion of the wavefunctions to $l=1000$, and 
all terms in the expansion of $V$ leading to an exchange of angular momentum $\Delta l > 15$ between the electrons due to the Coulomb interaction  are neglected. 
With these choices the numerical expense of calculating one value of $\Gamma_{M}$ still takes up to 20 hours 
on a 16 core Intel E5-2640v3  compute node. We estimate that the numerical uncertainty in $\Gamma_{M}$ due to these
approximations is approximately $10\%$ for initial states with $\delta q/Q >10^{-4}$, while we achieve full convergence for states with $\delta q/Q \le 10^{-4}$. 
\section{Selection of random states \label{selection}}
The two-atom states in Eq.~(\ref{state}) are independent-atom states and thus not eigenstates of the total Hamiltonian $H$. For sufficiently small 
atomic separations $R$, the interaction $V$ couples many states $\ket{\psi_M}$ with different quantum numbers~\cite{schwettmann:07,cabral:11}. 
Here we are not interested in the quantum dynamics of a particular initial state, 
but the aim is to characterize the autoionization rates of a large variety of different states $\ket{\psi_M}$. To this end, 
we calculate the autoionization rate $\Gamma_M$ for four sets $\mc{S}_i$ of  randomly chosen states that we select as follows. 
We consider  four non-overlapping energy intervals  that are centered around  the energies $\mc{E}_1$, $\mc{E}_2$, $\mc{E}_3$ and $\mc{E}_4$ of the $ndnd$ states  
with $n=14,16,18$ and $20$, respectively. We then find all two-atom manifolds $n_A l_A n_B l_B$ with $E_{n_A l_A}\ge E_{n_B l_B}$ and within an energy interval 
of $\pm 5\%$ around $\mc{E}_i$ and denote this set of manifolds by $\mc{M}_i$. We only retain
those  manifolds in $\mc{M}_i$ with orbital angular momentum $l_A,l_B\le4$.  For each set of manifolds $\mc{M}_i$ we choose one atomic separation $R_i$ such 
that most quantum numbers within $\mc{M}_i$ give rise to outer turning points $R_A$ and $R_B$ with 
\begin{align}
 0.8 \le \frac{R_i}{R_A+R_B} \le 1.4 \,. 
 \label{cond}
\end{align}
We find that this regime of weakly overlapping electron clouds can be adjusted by choosing 
$R_{1}=700 a_0$, $R_{2}=900 a_0$, $R_{3}=1200 a_0$ and $R_{4}=1500 a_0$,  and all manifolds  within $\mc{M}_i$ that do not obey Eq.~(\ref{cond}) are disregarded. 
After this pre-selection process each set $\mc{M}_i$ typically contains several hundred manifolds, and we randomly select 100 manifolds in $\mc{M}_i$ that 
form the set $\mc{S}_i$. Since the total magnetic quantum number is conserved  by the interaction Hamiltonian $V$, we confine our analysis to the $M=0$ subspace and assign 
each $n_Al_A$ manifold in $\mc{S}_i$ a random magnetic quantum number $m_A$ with $m_B=-m_A$. The distribution of the chosen states with respect to the 
principal quantum numbers $n_A$ and $n_B$ is shown in Fig.~\ref{fig2} for all four sets. It follows that each set contains a broad distribution of principal quantum numbers where 
the variation in both $n_A$ and $n_B$ is larger than five.
%
%
\begin{figure}[t!]
\begin{center}
\includegraphics[width=7.5cm]{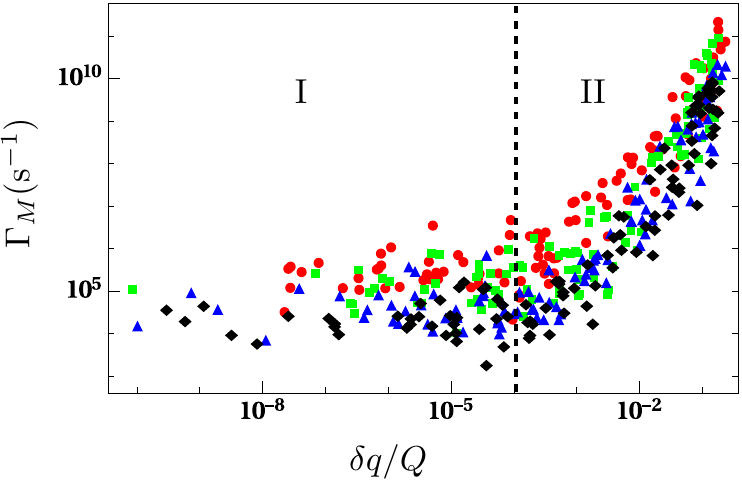}
\end{center}
\caption{\label{fig3}
(Color online)
Log-log plot of the autoionization  rates $\Gamma_M$ of the randomly chosen states $\mc{S}_i$ as a function of  $\delta q/Q$.  
Red dots correspond to $\mc{S}_1$, 
green squares show  $\mc{S}_2$, 
blue triangles are for  $\mc{S}_3$ 
and black diamonds correspond to  $\mc{S}_4$. 
The dashed line at $\delta q/Q=10^{-4}$ separates regions I and II where the autoionization  rates behave qualitatively different as a function of the overlap. 
}
\end{figure}
%
\section{Results and Discussion \label{res}}
The results for the autoionization  rate $\Gamma_M$ of the randomly chosen states in all sets $\mc{S}_i$ are shown in Fig.~\ref{fig3} as a function of the overlap $\delta q/Q$, see Eq.~(\ref{deltaq}).
There are two qualitatively different regions I and II divided by the dashed line at $\delta q/Q = 10^{-4}$. In region I 
the decay rates  appear to be  independent of the overlap. On the contrary, the decay rates increase sharply with $\delta q/Q$ in region II.  
In order to understand the physical reason for these two regions we perform reference calculations 
where we replace the interaction Hamiltonian $V$ in Eq.~(\ref{V}) 
by its multipole expansion $V_{\text{ME}}$~\cite{flannery:05} including dipole-dipole, dipole-quadrupole and quadrupole-quadrupole interactions. We find that the autoionization  rates calculated 
with $V_{\text{ME}}$  differ by at most $10\%$ from the values obtained with $V$ for all states with $\delta q/Q\le 10^{-4}$. We thus conclude that the multipole expansion of the 
interaction Hamiltonian holds if the overlap between the Rydberg orbitals is less than $10^{-4}$. 
On the other hand, the results obtained by $V_{\text{ME}}$ and $V$ differ greatly in region II. While all autoionization rates obtained by $V_{\text{ME}}$ are smaller than $3\times 10^7\text{s}^{-1}$, 
those calculated with $V$ can be  several orders of magnitude larger for $\delta q/Q\ge 10^{-2}$. This dramatic increase in the autoionization  rates with the 
overlap is consistent with the findings in~\cite{averbukh:04}. It can be explained physically by noting that  the full interaction Hamiltonian $V$ allows 
for direct electron-electron interactions in the region where the charge densities overlap, whereas the leading term in $V_{\text{ME}}$ is the dipole-dipole interaction. 
%

\begin{figure}[t!]
\begin{center}
\includegraphics[width=7.5cm]{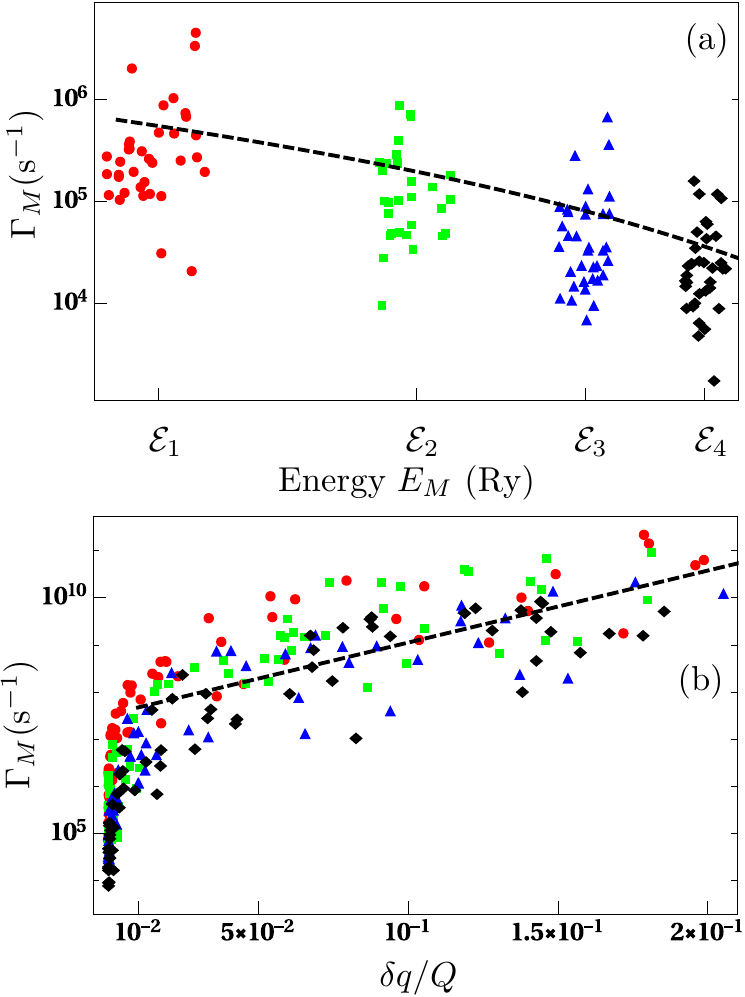}
\end{center}
\caption{\label{fig4}
(Color online)
(a) Log-linear plot of the autoionization rates $\Gamma_M$ of the randomly chosen states $\mc{S}_i$ in region I as a function of energy $E_M$ of the initial state.
 $\mc{E}_i$ is the central energy of the interval corresponding to set $\mc{S}_i$ (see Sec.~\ref{selection}), and Ry is the Rydberg constant. The dashed line 
 interpolates the four mean decay rates $\mean{\Gamma_M}_i$  [see Eq.~(\ref{qfit})]. 
(b) Log-linear plot of the autoionization  rates $\Gamma_M$ of the randomly chosen states $\mc{S}_i$ in region II as a function of  $\delta q/Q$. The dashed 
line is an exponential fit to the decay rates $\Gamma_M$ in all data sets $\mc{S}_i$ with $\delta q/Q\ge 10^{-2}$ [see Eq.~(\ref{fit})]. 
In (a) and (b), 
red dots correspond to $\mc{S}_1$, 
green squares show  $\mc{S}_2$, 
blue triangles are for  $\mc{S}_3$ 
and black diamonds correspond to  $\mc{S}_4$. 
}
\end{figure}
%
%
In the following we  analyse the autoionization  rates in regions I and II  in more detail. 
First, we focus on  region I and plot all  autoionization  rates $\Gamma_M$ with $\delta q/Q\le 10^{-4}$  as a function of the energy  $E_M$ of the initial state $\ket{\psi_M}$ 
as shown in Fig.~\ref{fig4}(a). Within each set $\mc{S}_i$, the autoionization rates show no evident energy dependence. The spread in $\Gamma_M$ is roughly the same for each set $\mc{S}_i$ and 
spans  about two orders of magnitude. However, the lower and upper bounds of each set $\mc{S}_i$ depend on energy such that the mean decay rates  $\mean{\Gamma_M}_i$ become gradually smaller 
by moving from set $\mc{S}_1$ to $\mc{S}_4$, where $\mean{\Gamma_M}_i$ is obtained by averaging over all decay rates $\Gamma_M$ in $\mc{S}_i$ with $\delta q/Q\le 10^{-4}$. 
This is illustrated by the dashed line in Fig.~\ref{fig4}(a)  interpolating the four mean decay rates $\mean{\Gamma_M}_i$, 
\begin{align}
 \mean{\Gamma_M} = \kappa (|E_M|/\text{Ry})^{\gamma},
 \label{qfit}
\end{align}
where $\kappa=2.51\times 10^{12}\text{s}^{-1}$, $\gamma=3.52$ and Ry is the Rydberg constant. 
The dominant contribution to the autoionization  rate in the multipole regime is the dipole-dipole interaction term such that $\Gamma_M^b\propto  d_b^2 d_c^2 R^{-6}$~\cite{amthor:09}, 
 where $d_b$ is the dipole matrix element between $\ket{\psi_B}$ and $\ket{\psi_b}$, and $d_c$ is the dipole matrix element 
between $\ket{\psi_A}$ and a Coulomb wave. The average principal quantum number $\bar{n}$ of the involved Rydberg states increases from set $\mc{S}_1$ to 
set $\mc{S}_4$, and hence we expect the involved dipole matrix elements to increase on average with $\bar{n}^2$~\cite{walker:08}. However, 
all states in a given set $\mc{S}_i$ are evaluated at a given atomic separation $R_i$ [see Sec.~\ref{selection}] with $R_i \approx R_A+R_B\propto \bar{n}^2$ [see Eq.~(\ref{cond})]. 
It follows that $\Gamma_M^b \propto \bar{n}^{-4}$, and hence we expect the full autoionization  rate to decrease with increasing energy of the two-atom state. 
On the other hand, the large spread in $\Gamma_M$ within each set $\mc{S}_i$ can be explained with the strong dependence of the  transition dipole matrix elements on the quantum numbers of the initial 
and bound states.

Second, we  analyze the steep increase of $\Gamma_M$ in region II. A log-linear plot of 
the autoionization  rates $\Gamma_M$  in region II is shown in Fig.~\ref{fig4}(b) as a function of  $\delta q/Q$. We find  that 
$\Gamma_M$ increases approximately exponentially for $\delta q/Q\ge 10^{-2}$. This is illustrated by the dashed line in Fig.~\ref{fig4}(b) given by
\begin{align}
\Gamma_M=  \Gamma_0  10^{\alpha x} ,
\label{fit}
\end{align}
where the parameters $\Gamma_0=3.31\times 10^7\text{s}^{-1}$ and $\alpha = 15.28$ are obtained by fitting the data points from all sets $\mc{S}_i$ with  $\delta q/Q\ge 10^{-2}$ to Eq.~(\ref{fit}). 
The spread of the decay rates around the dashed line is roughly 
three orders of magnitude for $\delta q/Q\le 0.15$, and reduces to two orders of magnitude for $\delta q/Q >  0.15$. 
In particular, the autoionization  rates are apparently independent of the energy of the initial state if the overlap exceeds several percent. 

\begin{figure}[t!]
\begin{center}
\includegraphics[width=7.5cm]{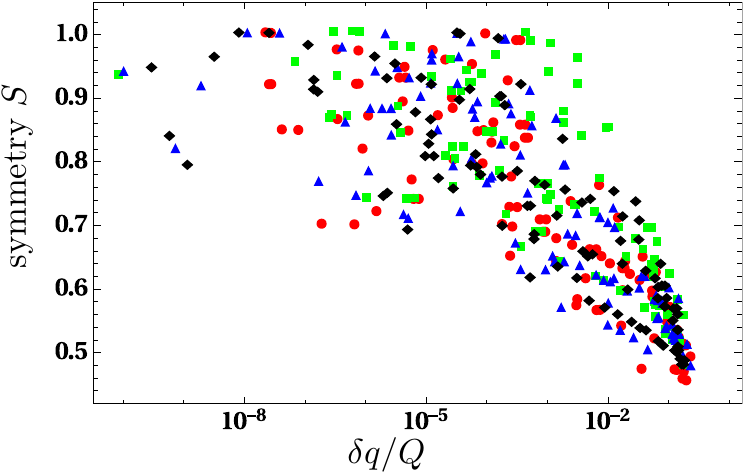}
\end{center}
\caption{\label{fig6}
(Color online)
Log-linear plot of the symmetry $S$ [see Eq.~(\ref{symm})]  of the randomly chosen states $\mc{S}_i$ as a function of  $\delta q/Q$.  
Red dots correspond to $\mc{S}_1$, 
green squares show  $\mc{S}_2$, 
blue triangles are for  $\mc{S}_3$ 
and black diamonds correspond to  $\mc{S}_4$. 
}
\end{figure}

Finally, we note that the overlap of a given state $\ket{\psi_M}$ is correlated with the symmetry of the energy distribution between 
the two atoms. More specifically, we consider the symmetry parameter 
\begin{align}
 S = 2 \frac{E_{n_A l_A}}{E_M} ,
 \label{symm}
\end{align}
where $E_{n_A l_A}$ is the independent-atom energy of state $\ket{\psi_A}$ and $E_M$ is defined in Eq.~(\ref{energy}). 
A value of $S=1$ corresponds to a completely symmetric distribution of energy $E_M$ between atoms $A$ and $B$, and $S$ decreases monotonically 
with reduced symmetry. 
Figure~\ref{fig6} shows a log-linear plot of $S$ for all sets of states $\mc{S}_i$ as a 
function of $\delta q/Q$, demonstrating that  symmetry and  overlap are clearly correlated.  
%
%
This result is relevant for systems similar to the  experimental setup reported in~\cite{tanner:08}, where a gas of cold atoms was excited to 
$ndnd$ states by short laser pulses. This initial  state is perfectly symmetric with $S=1$. However, the interatomic distance of some of 
the atom pairs in the gas will be so small that the dipole-dipole interaction 
couples the initial state to  near-resonant two-atom states with $S<1$. It follows that even if the autoionization rate of the initial $ndnd$ state 
is small for atomic pairs with $\delta q/Q<10^{-4}$, some of the two-atom states involved in the dipole-dipole cascades 
may autoionize much faster because they have $S<1$ and hence their overlap can be significantly larger than for the initial state. 
A more quantitative analysis of this point can be achieved by a simulation of the full quantum dynamics starting from an experimentally achievable initial state 
and including all coherent couplings between two-atom states and their autoionization rates. Such an investigation would be an interesting prospect for future studies. 
\section{Conclusion \label{conclusion}}
In this paper we present quantum mechanical calculations for autoionization rates of two nearby  Rydberg atoms. We consider 
sets of randomly chosen two-atom states and calculate the autoionization  rates in lowest order perturbation theory.  
Since the electron clouds overlap only slightly, we neglect exchange corrections to the autoionization  rate. 
We find that the autoionization rates can be classified via the charge overlap between the two states. 
If the overlap is less than $10^{-4}$, the multipole expansion of the interaction Hamiltonian holds and  the autoionization  rates are 
relatively small. In particular, they decrease on average with increasing energy of the two-atom state and can be smaller or comparable 
to  dipole transition rates between near-resonant two-atom states. It follows that the quantum dynamics in this regime will exhibit a rich 
interplay between coherent transitions and autoionization. However, we find that the autoionization 
rates increase dramatically beyond the dipole-dipole regime where overlap effects become significant. Our results show that this regime 
begins where the overlap exceeds $10^{-4}$, and an approximately exponential increase sets in if the overlap is larger than $1\%$. 
Our calculations were carried out for the specific example of Rubidium atoms. However, our classification of the autoionization rates in terms 
of the charge overlap makes no reference to the quantum numbers of the initial states or specific properties of Rubidium atoms. We thus expect that our findings hold 
for other alkali-metal atoms as well. 
While we had to restrict our calculations to relatively small principal quantum numbers due to technical reasons, we anticipate that 
qualitatively similar results should hold for higher principal quantum numbers as well. Extending our current calculations to this regime 
is subject to further investigation. Other possible extensions of our work include the calculation of the correct two-electron eigenstates via full configuration interaction methods~\cite{olsen:88}, 
and the application of the complex rotation method~\cite{ho:83} in order to find the energies and widths of the two-electron resonances.
In summary, Rydberg atoms with slightly overlapping electron clouds offer fascinating  possibilities for future
theoretical and experimental studies at the boundary between ultracold atom and molecular physics.
In particular, ultrafast pump-probe laser techniques~\cite{takei:15} allow one to resolve processes that are much faster than the autoionization 
rate even if the electron clouds overlap by a few percent. 
In this way autoionization and coherent processes in  correlated Rydberg electron clouds could be measured with
unprecedented temporal and spatial resolution. Such experiments would represent a paradigm shift from mimicking electron-electron
interactions with ultracold atoms~\cite{joerdens:12,palmer:06,cooper:13,bloch:08,simon:11,sanner:12}  to actually realizing them. 
\begin{acknowledgments}
We thank the National Research Foundation and the Ministry of Education of
Singapore for support.
The authors would like to acknowledge the use of the University of Oxford Advanced Research Computing (ARC) facility in 
carrying out this work (http://dx.doi.org/10.5281/zenodo.22558).
The research leading to these results has received
funding from the European Research Council under the European Unionʼs Seventh
Framework Programme (FP7/2007-2013)/ERC Grant Agreement no. 319286 Q-MAC.
\end{acknowledgments}
\appendix
\section{Evaluation of the Coulomb matrix element \label{matrixE}}
Here we outline the evaluation of the matrix element 
\begin{align}
 M=\bra{\psi_{l_k m_k}^{E_k},\psi_b}V\ket{\psi_A,\psi_B}
 \label{Mel}
\end{align}
entering the autoionization rate in  Eq.~(\ref{auger1}). The operator $V$ in Eq.~(\ref{V}) is a sum of Coulomb interactions $1/|\mf{r}-\mf{r}'|$ 
which whe expand as 
\begin{align}
 \frac{1}{|\mf{r}-\mf{r}'|}=
\sum\limits_{l=0}^{\infty}\sum\limits_{m=-l}^{l}\frac{4\pi}{2 l +1}
\frac{r_<^l}{r_>^{l+1}}Y_l^{m^*}(\theta',\phi')Y_l^m(\theta,\phi),
\label{laplace_exp}
\end{align}
where $r_<=\text{min}(r,r')$ and $r_>=\text{max}(r,r')$. 
We truncate the sum over angular momenta $l$ in Eq.~(\ref{laplace_exp})  
and omit all terms with  $l>15$. Here $l$ corresponds to $\Delta l$ in the main text and determines the amount of angular momentum 
that the Coulomb interaction can transfer between the electrons. 
In order to evaluate the matrix element $M$ we expand all wavefunctions in terms of spherical harmonics~\cite{tannoudji:qm}. 
Since we place atom $A$ at the origin, the expansion of $\ket{\psi_A}$ and $\ket{\psi_{lm}^E}$  comprise only  a single term, 
\begin{subequations}
\label{expansion1}
\begin{align}
&\psi_A(\mf{r})=  R_{n_A l_A}(r) Y_{l_A}^{m_A}(\theta,\phi),  \\
&\psi_{lm}^E(\mf{r})=  C_{El}(r) Y_{l}^{m}(\theta,\phi).
\end{align}
\end{subequations}
The wavefunctions $\psi_B(\mf{r})$ and $\psi_b(\mf{r})$ in Eq.~(\ref{Mel}) are centered at  atom $B$. They are both of the form 
\begin{align}
 \psi_{\beta}(\mf{r})=\psi_{n_{\beta} l_{\beta} m_{\beta}}(\mf{r}-\mf{R})
\end{align}
 with $\psi_{n_{\beta} l_{\beta} m_{\beta}}(\mf{r})=R_{n_{\beta} l_{\beta}}(r)Y_{l_{\beta}}^{m_{\beta}}(\theta,\phi)$ and $\beta\in\{b,B\}$. 
Since $\mf{R}$ is different from zero we make a general ansatz for  the expansion of $\psi_{\beta}(\mf{r})$ in terms of spherical harmonics, 
\begin{align}
\psi_{\beta}(r,\theta,\phi) =  \sum\limits_{l_q=0}^{L_{\text{max}}}\sum\limits_{m_q=-l_q}^{l_q}Q_{l_q}^{m_q}(r) Y_{l_q}^{m_q}(\theta,\phi), 
\label{expansion2}
\end{align}
where we expressed $\mf{r}$ in terms of spherical coordinates $\mf{r}(r,\theta,\phi)$. 
We set $L_{\text{max}}=1000$, and the function $Q_{l_q}^{m_q}(r)$ can be found using the orthonormality of  $Y_{l_q}^{m_q}$,
\begin{align}
 Q_{l_q}^{m_q}(r) = \int\limits\text{d}\theta \text{d}\phi \sin\theta \psi_{\beta}(r,\theta,\phi)Y_{l_q}^{m_q^*}(\theta,\phi) \,. \label{spherical}
\end{align}
We represent all radial functions on a grid with up to  14000 points. The integration region in Eq.~(\ref{spherical}) is restricted to the 
solid angle where $\psi_{\beta}(\mf{r})$ takes on non-negligible values, and the integral is carried out using the trapezoidal rule~\cite{ms}. 
With the expansions in Eqs.~(\ref{laplace_exp}),~(\ref{expansion1}) and~(\ref{expansion2}) the evaluation of the matrix element $M$ 
can be reduced to a double integral over the radial variables and the remaining
integrals reduce to Gaunt coefficients~\cite{tannoudji:qm2}.  
The radial integrals are evaluated  with the trapezoidal rule~\cite{ms}, and 
the Gaunt coefficients are defined as 
\begin{align}
 G_{l_1l_2l_3}^{m_1m_2m_3} = &\int\text{d}\theta \text{d}\phi \sin\theta  Y_{l_1}^{m_1}(\theta,\phi)Y_{l_2}^{m_2}(\theta,\phi)Y_{l_3}^{m_3}(\theta,\phi) \notag \\
 = & (-1)^{m_3}\sqrt{\frac{(2l_1+1)(2l_2+1)}{4\pi(2l_3+1)}} \notag \\
 & \braket{l_1,l_2;0,0}{l_3,0}\braket{l_1,l_2;m_1,m_2}{l_3,-m_3} . 
\end{align}
The evaluation of  Clebsch-Gordan coefficients~\cite{tannoudji:qm2}  $\braket{l_1,l_2;m_1,m_2}{l_3,m_3}$ 
involves the calculation of factorials which can be numerically unstable for large values of $l_1,\,l_2$ and $l_3$ if floating point numbers are used. In order 
to circumvent this problem, we generate a library of all non-zero Gaunt coefficients with $l_1,\,l_2\le 1000$ and $l_3\le 15$ with the 
software packet MATHEMATICA~\cite{MM}. The calculation of $\Gamma_A$ is implemented in MATLAB~\cite{MA}. 
\end{document}